\documentclass{elsart4-1}

\usepackage{graphicx}
\usepackage{units}

\usepackage{amssymb}

\usepackage[english,francais]{babel}

\newtheorem{e-proposition}[theorem]{Proposition}

\newtheorem{e-definition}[theorem]{Definition\rm}

\renewcommand{\theequation}{\arabic{equation}}
\def\og{\leavevmode\raise.3ex\hbox{$\scriptscriptstyle\langle\!\langle$~}}
\def\fg{\leavevmode\raise.3ex\hbox{~$\!\scriptscriptstyle\,\rangle\!\rangle$}}

\begin{document}

\begin{frontmatter}


\selectlanguage{english}
\title{Nucleation of magnetisation reversal, from nanoparticles to bulk materials}


\selectlanguage{english}
\author[JV]{Jan Vogel},
\ead{jan.vogel@grenoble.cnrs.fr}
\author[JM]{J\'er\^{o}me Moritz},
\ead{jerome.moritz@cea.fr}
\author[JV]{Olivier Fruchart}
\ead{olivier.fruchart@grenoble.cnrs.fr}

\address[JV]{Institut N\'{e}el, CNRS, 25 rue des Martyrs,
B.P.166, F-38042 Grenoble Cedex 9, France}
\address[JM]{SPINTEC (URA 2512 CNRS/CEA), CEA/Grenoble, 17 Rue des Martyrs, F-38054 Grenoble Cedex 9, France}



\begin{abstract}

We review models for the nucleation of magnetisation reversal, i.e.
the formation of a region of reversed magnetisation in an initially
magnetically saturated system. For small particles models for
collective reversal, either uniform (Stoner-Wohlfarth model) or
non-uniform like curling, provide good agreement between theory and
experiment. For microscopic objects and thin films, we consider two
models, uniform (Stoner-Wohlfarth) reversal inside a nucleation
volume and a droplet model, where the free energy of an inverse
bubble is calculated taking into account volume energy (Zeeman
energy) and surface tension (domain wall energy). In macroscopic
systems, inhomogeneities in magnetic properties cause a distribution
of energy barriers for nucleation, which strongly influences effects
of temperature and applied field on magnetisation reversal. For
these systems, macroscopic material parameters like exchange
interaction, spontaneous magnetisation and magnetic anisotropy can
give an indication of the magnetic coercivity, but exact values for
nucleation fields are in general hard to predict.

{\it To cite this article: J. Vogel, J. Moritz, O. Fruchart, C. R.
Physique 7, 977 (2006).}

\vskip 0.5\baselineskip

\selectlanguage{francais} \noindent{\bf R\'esum\'e} \vskip
0.5\baselineskip \noindent {\bf Nucl\'eation du renversement de
l'aimantation, de nanoparticules aux syst\`emes macroscopiques.}
Nous passons en revue diff\'{e}rents mod\`{e}les traitant de la
nucl\'{e}ation dans des syst\`{e}mes magn\'{e}tiques. La
nucl\'{e}ation est l'\'{e}tape qui initie le renversement de
l'aimantation dans des objets magn\'{e}tiques pr\'{e}alablement
satur\'{e}s en champ. Pour des particules d'extension spatiale
r\'{e}duite -typiquement inf\'{e}rieure \`{a} la largeur de paroi de
domaines magn\'{e}tiques- les mod\`{e}les de renversement collectif,
homog\`{e}ne (mod\`{e}le de Stoner-Wohlfarth) ou par exemple de type
'curling' donnent un bon accord avec l'exp\'{e}rience. Pour des
objets microscopiques et des couches minces, nous discutons deux
approches, l'une supposant le renversement homog\`{e}ne de
l'aimantation dans un volume d'activation de taille r\'{e}duite,
l'autre fond\'{e}e sur l'\'{e}nergie de formation d'une gouttelette.
Cette derni\`{e}re est calcul\'{e}e en tenant compte de
l'\'{e}nergie de volume (terme Zeeman) et de surface (\'{e}nergie de
paroi). Les deux approches sont utilis\'{e}es pour d\'{e}terminer
des volumes d'activation ou ajuster des courbes exp\'{e}rimentales.
Il semblerait que le mod\`{e}le de gouttelette soit plus pertinent
pour expliquer les variations thermiques des champs de renversement
ou leur comportement dynamique. Dans les syst\`{e}mes
macroscopiques, les inhomog\'{e}n\'{e}it\'{e}s du mat\'{e}riau
donnent lieu \`{a} des distributions d'\'{e}nergie de barri\`{e}re
qu'il est n\'{e}cessaire d'int\'{e}grer dans les calculs. Ainsi les
param\`{e}tres comme l'anisotropie ou l'\'{e}change peuvent donner
une indication de la coercitivit\'{e}, mais il est en
g\'{e}n\'{e}ral difficile de pr\'{e}voir sa valeur exacte.

{\it Pour citer cet article~: J. Vogel, J. Moritz, O. Fruchart, C.
R. Physique 7, 977 (2006).}

 \vskip 0.5\baselineskip \noindent{\small{\it
Mots-cl\'es~:} nucl\'eation~; renversement de l'aimantation~;
couches minces magn\'etiques~; activation thermique}}
\end{abstract}
\end{frontmatter}


\selectlanguage{english}
\section{Introduction}
\label{Intro} The spontaneous or magnetic field-induced reversal of
the magnetisation direction in materials and nanostructures has been
the subject of intensive studies since many decades. The use of
magnetic materials in applications ranging from compass needles to
electrical motors, truck brakes, cellular phones and personal
computers has triggered the search for materials with particular
properties concerning both their static and dynamic behaviour.
Materials used \textit{e.g.} in transformers need to switch their
magnetisation direction under very small values of applied magnetic
fields to minimise losses; these are so-called soft magnetic
materials. On the other hand, permanent magnets used in electrical
motors and generators need their magnetisation to be as stable as
possible against both magnetic fields and thermal effects; these are
so-called hard magnetic materials. Consider however information
written on magnetic storage media in the form of small magnetic
grains. In this case, a compromise must be found for the grains need
to be stable for years, but on the other hand their magnetisation
direction should respond quickly to moderate magnetic field values
to allow information to be written fast. Thus for each particular
case an understanding of how magnetisation reversal proceeds is
needed. Magnetisation reversal can take place in different ways,
depending on the size of the object and physical parameters like the
exchange interaction and magnetic anisotropy. In the atomic case, a
magnetic field along a direction other than the initial
magnetisation will cause a torque on the magnetisation given by
\textbf{M} $\times$ \textbf{H$_{eff}$}, where \textbf{H$_{eff}$} is
the local effective field. This torque will induce a precessional
motion of the magnetisation around the effective field. In
materials, damping will occur and eventually the magnetisation will
align with the effective field. This scheme is described by the
Landau-Lifshitz-Gilbert (LLG) equation \cite{LLG} of precession and
damping. This equation remains valid also for very small particles,
i.e. when the magnetic moments of the different atoms are strongly
coupled one to another by exchange interaction and act like a
so-called \textsl{macro-spin}. This simple reversal scheme is
referred to as coherent or uniform magnetisation reversal.

In nearly all materials the rotational symmetry is lost as a
consequence of the discrete crystalline structure and the spin-orbit
coupling, or due to dipolar interactions. This effect is called
magnetic anisotropy, which can be expressed as an angular-dependent
energy. In the presence of anisotropy, a field of finite size is
needed to reverse the magnetisation. For uniaxial anisotropy the
angular dependence of the uniform reversal has been derived by
Stoner and Wohlfarth \cite{StonerWohlfarth} (SW). These features are
treated in Section~\ref{sec-uniform}. For micrometer-sized particles
or dots the macrospin approximation is not exact anymore and local
deviations from uniform magnetisation usually exist. Interestingly,
notice that precession-like magnetisation reversal may still take
place at the micrometer scale \cite{Schumacher2003a}, but in that
case the non-uniform magnetisation has to be taken into account for
an accurate description of the magnetisation reversal
\cite{MiltatDynamicsbook}. This is discussed in
Section~\ref{sec-elongated}.

For macroscopic films (Sec. \ref{sec-films}) and bulk magnetic
materials (Sec.~{sec-bulk}), magnetisation reversal usually starts
at magnetic fields that are significantly lower than expected from
the theory of uniform reversal of the entire system; This is the
so-called Brown's paradox \cite{Brown45}). Rather, in these systems
magnetisation reversal takes place through a process of nucleation
of small reversed domains (usually on inhomogeneities or defects)
and a subsequent propagation of magnetic domain walls. Here we will
treat nucleation processes only. The propagation of domain walls,
once a stable nucleation center is formed, is not discussed. Over
the years, different models have been developed to take into account
the role of nucleation in non-uniform magnetisation reversal and to
explain experimentally obtained values of reversal fields and other
dynamic properties of magnetic materials and films. These basic
models will be reviewed in this paper, for small particles and dots
as well as for magnetic thin films. This overview will be
non-exhaustive, and given from an experimentalist point of view. In
some cases, a comparison with experimentally obtained results will
be made. The role of inhomogeneities in the samples themselves (like
defects and finite-size effects) on the nucleation of reversed
domains will be discussed. In the following sections magnetisation
reversal and nucleation is considered in systems of increasing
dimensionality and thus complexity, from zero-dimensional to
three-dimensional, \textit{i.e.} bulk.

\section{Uniform or 'coherent' reversal}
\label{sec-uniform}For very small particles, the exchange
interaction is dominating and induces uniform magnetisation. The
upper critical size for uniform magnetisation depends on the
detailed shape of the particle and on the material parameters
\cite{Brown57,Frei57,Fruchart05}. For particles smaller than this
critical size, magnetisation reversal will be uniform and nucleation
processes as defined above can not take place. Even though uniform
magnetisation reversal is not the subject of this paper, it is
briefly mentioned here since uniform reversal is also the base of
models for nucleation of reversed domains in macroscopic materials.
In that case, a so-called \textsl{activation volume} is considered,
in which the magnetisation is assumed to reverse coherently, and the
energy for this reversal may be calculated assuming an independent
volume or taking into account the interaction of the reversed volume
with the matrix.

The energy of a uniformly magnetised particle of volume $V$ in an
external magnetic field \textbf{H} depends on the spontaneous
magnetisation \textit{M$_S$} and on the orientation \textbf{r} of
the magnetisation with respect to the external field and with
respect to the anisotropy axes of the particle \cite{Bonet99}:

\begin{equation}\label{Particle}
    E(r,H) = E_K(r) - \mu_0M_SV\bf r \bullet \bf H
\end{equation}

\noindent where E$_K$(\textbf{r}) is the energy caused by the
magnetic anisotropy of the particle and the second term is the
Zeeman energy. Stoner and Wohlfarth \cite{StonerWohlfarth}
considered the simple case of magnetisation reversal  in a single
domain particle with uniaxial magnetic anisotropy and a positive
anisotropy constant \textit{K}. In that case, \textit{E$_K$(r) = -
KVcos$^2\theta$}, where $\theta$ is the angle between the direction
of magnetisation and the easy axis. A two-dimensional plot of the
switching field as a function of $\theta$ results in the well-known
Stoner-Wohlfarth astroid \cite{Slonczewski56}. The SW model was
later extended to macro-spin particles with an arbitrary
three-dimensional anisotropy by Thiaville \cite{Thiaville00}, with
an experimental demonstration given by Bonet \textit{et al.}
\cite{Bonet99}. In all these models, the angular dependence of the
reversal field of the particle is derived by calculating the field
for which the energy barrier $\Delta$E(H) that has to be overcome to
reach the minimum energy state vanishes. This means that thermal
effects that may help to overcome this barrier are not taken into
account. Thermally-assisted reversal has first been considered by
N\'{e}el \cite{Neel49} and Brown \cite{Brown63}. In the so-called
N\'{e}el-Brown model, the probability that the magnetisation of a
particle has switched after time \textit{t} is given by P(t) = 1 -
e$^{-t/\tau}$, where the time constant $\tau$ can be expressed by an
Arrhenius law of the form : $\tau$(T,H) = $\tau_0$ e$^{\Delta
E(H)/kT}$. $\tau_0$ is an attempt frequency of the order of
$10^{-10}-10^{-9}$ seconds. An experimental verification of the
N\'{e}el-Brown model was given by Wernsdorfer \textit{et al.}
\cite{Wernsdorfer97}.

\begin{figure}[b]
\includegraphics*[bb= 116 516 480 634]{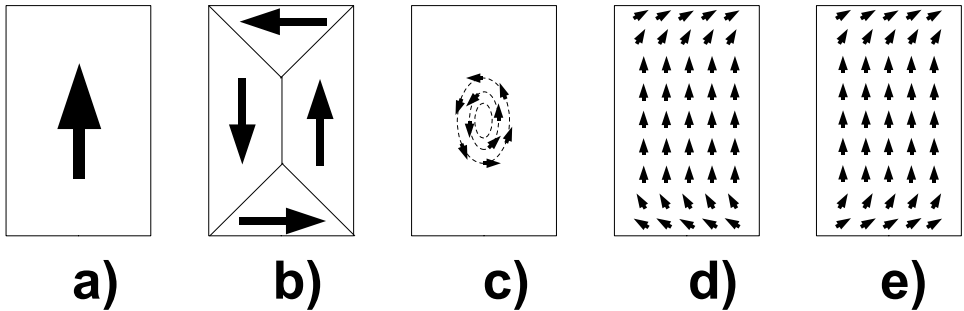}
\caption{\label{buckling} Some possible magnetisation configurations
of a rectangular platelet {\bf a)} collinear, single domain
structure {\bf b)} flux-closure domain state {\bf c)} curling of
magnetisation ('vortex') {\bf d)} C-state {\bf e)} S-state}
\end{figure}

The aforementioned models are valid for strictly uniform
magnetisation reversal. Usually, the reversal field can be lower
when the magnetisation is not uniform during the reversal. Brown
\cite{Brown57} and Frei \textit{et al.} \cite{Frei57} calculated the
size above which the magnetisation of a particle will not be uniform
anymore. Frei \textit{et al.} \cite{Frei57} calculated that for a
prolate ellipsoid the critical size shows little dependence on
magnetocrystalline anisotropy and exact shape and is approximately
equal to \textit{A$^{1/2}$/M$_S$}, where $A$ is the exchange
constant. For slightly larger particles, the magnetisation will stop
being uniform, and can arrange according to different spatial
distributions. For a flat, rectangular particle, for example,
micromagnetic simulations and models have revealed several possible
near-single domain configurations \cite{Chui97,Cowburn98} and some
of them are shown in Fig.~\ref{buckling}.

It is generally assumed that the SW model can be adapted to the case
of nucleation in thin films as far as magnetic moments rotate at
unison inside an activation volume. The barrier height $\Delta$E$_n$
to nucleation is then written under the form:

\begin{equation}\label{SWnucl}
    \Delta E_n = E_n (1 - \frac{H}{H_n})^\alpha
\end{equation}

\noindent where \textit{H$_n$} and \textit{E$_n$} are nucleation
field and nucleation energy, respectively. The value of $\alpha$
varies with the direction of the applied field with respect to the
easy axis. Victora has derived a \textit{H$^{3/2}$} scaling
relationship between \textit{H} and \textit{$\Delta$E} for small
values of \textit{$\Delta E_n$} \cite{Victora89}. N\'eel has
predicted earlier a quadratic dependence for the special case of
high symmetry \cite{Neel49a}, for any height of the barrier. Some
authors use an exponent equal to 1 to describe domain nucleation and
domain wall propagation \cite{Kirilyuk93}. More generally, we can
consider $1 \leq \alpha \leq 2$.

\section{Nucleation in elongated particles}
\label{sec-elongated} Here we consider the case of one-dimensional
systems. This can be realized in particles with a width below but a
length above about twice the domain wall width. In this case,
magnetisation switching can be initiated by reversal inside some
\textit{nucleation volume} much smaller than the total volume of the
particle. This non-uniform scheme of magnetisation reversal takes
place because the energy barriers associated with magnetisation
reversal are much lower than for uniform reversal. This tendency
will be ever more valid for increasing dimensionality, as described
in the following sections. In these particles, the main contribution
to the magnetic anisotropy is given by the shape anisotropy,
favoring a magnetisation parallel to the long axis. Braun
\cite{Braun93} developed a model for thermally activated
magnetisation reversal in this kind of particles. Considering
thermally assisted reversal in an activation volume much smaller
than the total particle volume, taking into account the exchange
interaction at the boundary of the volume, the author found an
energy barrier for reversal proportional to the cross-sectional area
of the particle. Qualitative agreement was found with
room-temperature measurements on single elongated particles by
Lederman \textit{et al.} \cite{Lederman95}. Wernsdorfer \textit{et
al.} \cite{Wernsdorfer96} performed low temperature measurements on
single polycrystalline Ni wires with lengths of some micrometers and
diameters between 40 and 100 nm. Their measurements gave evidence of
nucleation of magnetisation reversal in an activation volume of the
order of (20nm)$^3$. For the smallest diameters, they found that the
data could be fitted using a single energy barrier and that the
reversal process could be described by an Arrhenius law. For larger
diameters (75-100 nm), nucleation occurred at several values of the
applied field close to each other, and close to the value expected
for a curling mode \cite{Brown57,Frei57} of reversal. A statistic
analysis of the switching times showed that they could be fitted
using a stretched exponential \textit{P(t) = e$^{-(t/\tau)\beta}$}.
For the smallest wires $\beta$ was close to 1, as expected from the
N\'{e}el-Brown model. For wires with a diameter between 75 and 100
nm, values of $\beta$ between 0.1 and 0.5 were found, indicating a
distribution of energy barriers for nucleation.

\section{Nucleation in continuous and microstructured thin films}
\label{sec-films} In 2-dimensional magnetic films or single-domain
microstructures with in-plane dimensions considerably larger than
the domain wall width, the reversal of magnetisation is in general
initiated by nucleation. As soon as the effective field reaches some
critical value (the \textit{nucleation field}) some nuclei appear,
the number of nuclei depending on several parameters (see
sub-section \ref{statistics}). At this point, the local moments
start to rotate and the magnetisation configuration distorts,
leading to bubbles of inverse magnetisation. These bubbles, with a
minimum volume usually called 'nucleation volume' or 'activation
volume', are separated from the other magnetic phase by domain
walls.  Nucleation is a process characterized by the overcoming of
an energy barrier as depicted before. The barrier height depends
intrinsically on material constants like anisotropy, magnetisation
or exchange energy and decreases when the applied field increases.
In this section, we will describe the different activation laws
which govern nucleation in thin films and explain how experimental
data can corroborate the theoretical facts. Finally we will focus on
the structural defects or other physical causes, which facilitate
the appearance of inverse nuclei.

After nucleation, complete reversal is achieved by propagation of
these domain walls through the sample, or the particle, under the
field pressure. The magnetisation reversal properties of a material
are determined by a combination of the properties of nucleation and
domain wall propagation, with their associated energy barriers. The
properties of domain wall propagation are, however, not the subject
of this paper (see, for example, Ref.~\cite{Ferre02})

\subsection{Energy barriers}
There are two main ways to estimate the energy barriers to the
nucleation. One is to assume that uniform reversal takes place
inside the activation volume during nucleation
\cite{StonerWohlfarth}. The other is more macroscopic, considering
the nucleation volume as a uniformly reversed magnetic domain, and
its boundary as a domain wall. This so-called droplet theory is
based on the competition between volume energy and surface tension
\cite{Barbara94}.

\subsubsection{Coherent rotation inside the nucleation volume}
In the SW model of coherent rotation, {\it E$_n$} is the anisotropy
energy {\it K.V} and {\it H$_n$} the anisotropy field {\it H$_K$},
so the barrier height is cancelled when the field reaches {\it
H$_K$}. Equation \ref{SWnucl} implies that $\Delta E_n$ is a balance
between Zeeman and nucleation energies. Most often {\it E$_n$} is
assimilated to the product between an effective anisotropy {\it
K$_{eff}$} and a nucleation volume {\it V$_n$} and experimental data
can be fitted to estimate these. For instance Sharrock has shown
that switching volumes can be estimated by fitting the barrier
crossing and that some adjustments permit to link {\it V$_n$} to
volumes of particles used in granular magnetic recording media
\cite{Sharrock94}. More recently, it has been shown that for real
systems, {\it V$_n$} can be related to the volumes of grains only
for isolated identical grains, which is not really the case in
continuous magnetic media \cite{Yu99}. The nucleation volume depends
thus explicitly on $\Delta E_n$ and can be regarded as an average
unit volume of magnetic moments switching together. Thus $V_n$ can
be extrapolated from the formula \cite{Givord87}:

\begin{equation}\label{VnGivord}
    V_n = -\frac{\delta\Delta E_n}{\delta H}/M_S
\end{equation}

Nucleation volume and nucleation field are very sensitive to the
local environment. It has been simulated that dipolar coupling and
inter-granular exchange interactions act at short and long ranges on
the nucleation processes \cite{Schrefl93,Zhou02}. In the case of
flat dots with in-plane uniaxial anisotropy, Fruchart {\it et al.}
have proposed a model which predicts the barrier height to
nucleation when demagnetisation effects are taken into account as a
pinpoint torque located at the edge of the structures
\cite{Fruchart01}. They have shown that the energy barrier could be
quite different from the simple SW one. Models for magnetisation
reversal in flat dots made of soft magnetic material, based on the
description of the stability of local nucleation volumes, have also
been proposed \cite{Grimsditch01}. Notice that in all cases the
nucleation volume, and particularly the product $K_{eff}.V_n$,
determines the barrier height at zero field, and is therefore a
stability criterion to respect in the framework of magnetic
recording \cite{Weller99}. Ideally, we can consider that in
continuous layers with large and coupled grains, $V_n$ is about
$\delta_w^2.t$, with {\it t} the thickness of the film and
$\delta_w$ the domain wall width.

\subsubsection{Droplet model}
\begin{figure}[b]
\center \includegraphics*[bb= 160 441 356 586]{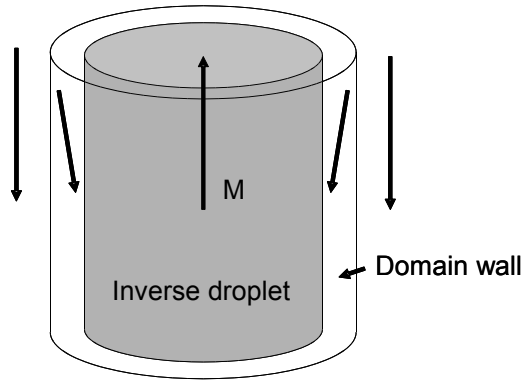}
\caption{\label{droplet} In the droplet model, the applied field
creates an inverse droplet into a saturated sample. The total energy
is a balance between the Zeeman energy (volume term) and the domain
wall energy (surface term).}
\end{figure}

Another way to model nucleation is to calculate the energy of an
inverse bubble by taking into account volume energy (Zeeman energy)
and surface tension (domain wall energy $\gamma$) (see
Fig.~\ref{droplet}). This was proposed by Barbara \cite{Barbara78}
to explain the thermal variation of the coercivity in a Dy$_3$Al$_2$
alloy, and has been used recently in the case of CoPtCr alloys
\cite{Kirby00} and Co/Pt thin films \cite{Moritz05}. It has also
been used for nucleation in a kinetic Ising model \cite{Richards95}.
The energy of a cylindrical droplet of radius $r$, appearing in a
magnetic thin film of thickness $t$ is:

\begin{equation}\label{Eq.droplet}
    F(r) = 2\pi r\gamma t - 2\pi r^2t\mu_0HM_S
\end{equation}

When the increase in Zeeman energy upon increasing size
counter-balances the loss in energy due to the increase of the
domain wall length, the droplet reaches a critical size given by
$\delta F(r)/\delta(r) = 0$ and the critical radius is $r_c =
\gamma/2M_S\mu_0H$. If $r < r_c$, droplets collapse. If $r > r_c$,
the droplet energy decreases as its radius grows, {\it i.e.} there
is propagation of the domain walls through the sample. The initial
increase of the radius is very slow, due to the surface tension
induced by the domain wall energy \cite{Skomski96,Fukumoto06}.

Using
the above formulas, it is possible to define the height of the
energy barrier for nucleation by:

\begin{equation}\label{CritBarr}
    \Delta E_n = F(r_c) = \frac{\pi\gamma^2t}{2M_S\mu_0H}
\end{equation}

A comparison of this expression with equation~\ref{SWnucl} shows
that taking into account the presence of domain walls and their
associated energy in 2-D films leads to a very different field
dependence of the nucleation barrier. Notably, for a field $H=0$ the
energy barrier will be infinite for continuous films, while it is
given by the magnetic anisotropy for small particles. An extension
of the droplet model to the case where the domain wall energy is not
homogeneous on the sample will be given in section~\ref{defects}.

Aharoni and Baltensberger have numerically calculated the total
magnetic energy of both cylindrical and spherical bubbles
\cite{Aharoni92}. They have shown that in the approximation of low
fields, the droplet and the micro-magnetic approaches converge.
Although it seems that the droplet model gives a satisfactory
explanation for the thermal variation of coercivity (see below), it
has not been extensively used to model nucleation.

Notice that the droplet model is relevant only for dimensions equal
or larger than two. Indeed in one dimension, like for the case of
elongated particles presented above, the increase of the region of
reverse magnetisation does not imply a significant increase of the
area of the domain wall. The formulas derived above are valid only
for $d=2$.

\begin{figure}[t]
\center \includegraphics*[bb= 183 324 409 605]{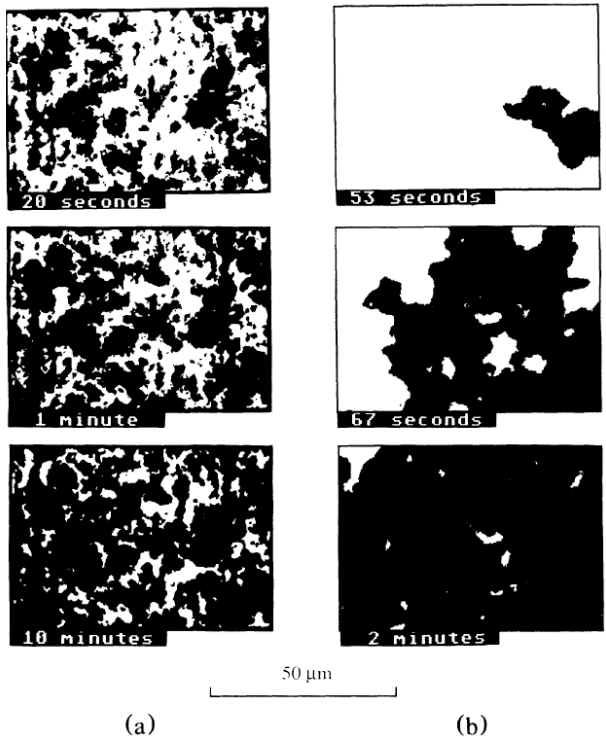}
\caption{\label{KerrCoPt} Time evolution of the magnetic domain
structures measured in Au/Co/Au sandwiches using Kerr microscopy for
two values of the applied magnetic field, H = 470 Oe {\bf (a)} and H
= 413 Oe {\bf (b)}. It is clearly seen that the number of inverse
bubbles and therefore the nucleation density depends on the field
strength. Reprinted figure with permission from J. Pommier, P.
Meyer, G.P\'enissard, J. Ferr\'e, P. Bruno, and D. Renard, Phys.
Rev. Lett. 65, 2054  (1990). Copyright 1990 by the American Physical
Society.}
\end{figure}

\subsection{Thermal and time effects}\label{statistics}
Overcoming the energy barrier for nucleation can be achieved using
an applied field to lower its value, but also temperature can help
by thermal activation. Like the Arrhenius-N\'eel law for the
reversal time $\tau$, the nucleation rate can be written as:

\begin{equation}\label{NuclRate}
    R = R_0 \exp(\frac{-\Delta E_n}{k_BT})
\end{equation}

Based on this nucleation rate, some authors have expressed the
reversed magnetisation area corresponding to the nucleation centers
and the propagation of the domain walls. These models are developed
based on Fatuzzo's theory of the relaxation of the polarization in
ferro-electrics \cite{Fatuzzo62}. In this model, the relaxation is
characterised by a single parameter $k$, that defines the relative
importance of nucleation and domain wall propagation processes in
the reversal. Labrune {\it et al.} have applied Fatuzzo's model to
the case of magnetic films submitted to a constant applied field and
showed that magnetic after-effect experiments performed on GdTbFe
thin films could be very well modeled this way \cite{Labrune89}.

\begin{figure}[b]
\center \includegraphics*[bb= 182 365 416 522]{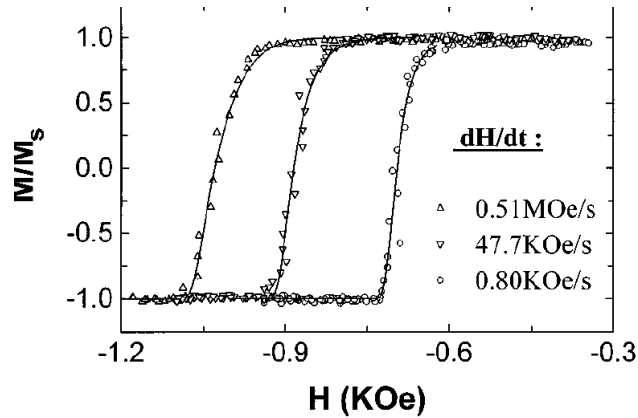}
\caption{\label{Raquet} Experimental hysteresis loops (open
triangles) and simulated ones (full lines) measured on Au/Co/Au
trilayers. A good agreement is found for low field sweep rates by
using a SW barrier form with $\alpha=1$. Reprinted figure with
permission from B. Raquet, R. Mamy, and J.C. Ousset, Phys. Rev. B.
54, 4128  (1996). Copyright 1996 by the American Physical Society.}
\end{figure}

Magnetic relaxation measurements coupled to Kerr microscopy have
been realized by Pommier {\it et al.} on films with a magnetisation
perpendicular to the film plane \cite{Pommier90} (see
Fig.~\ref{KerrCoPt}). They have found evidence that the reversal is
governed by two dominating processes, nucleation and propagation.
They have explained as well that local properties can generate a
distribution of nucleation fields. Later, Raquet {\it et al.} have
adapted Fatuzzo's model for varying magnetic fields with a constant
field sweep rate \cite{Raquet96}. They have applied their model to
the case of Au/Co/Au sandwiches and quantified the wall velocity,
the nucleation rate and the Barkhausen volume, the volume of
reversed magnetisation at each domain wall jump. They have used a
barrier height to nucleation following Eq.~(\ref{SWnucl}) with an
exponent equal to 1 and showed a good agreement between the
experimental measurements and their calculations for low applied
field sweep rates (see Fig.~\ref{Raquet}).

\begin{figure}[t]
\center \includegraphics*[bb= 198 390 411 556]{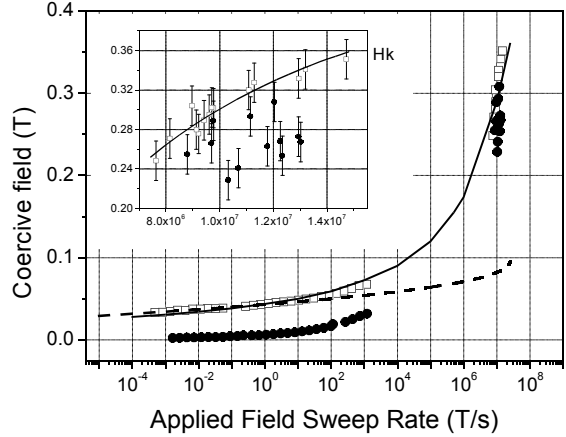}
\caption{\label{Moritz} Variation of the dynamical coercivity vs.
the applied field sweep rate measured in continuous (dots) and
patterned (squares) Co/Pt thin film multilayers. The continuous line
represents the calculations of {\it H$_C$} using the droplet model;
the dashed one is the calculated {\it H$_C$} deduced from the SW
model. The inset shows a zoom for field sweep rates around $10^7$
T/s (see Ref.~\cite{Moritz05}).}
\end{figure}

The possibility to pattern magnetic thin films has recently allowed
the study of a single nucleation center. Moritz {\it et al.} have
applied the droplet model to nucleation in perpendicular-to-plane
magnetised dots obtained by electron beam lithography
\cite{Moritz05}. The dots had sizes of about 100$\times$100nm$^2$,
allowing the hypothesis of only one nucleation center per dot. The
dynamical investigations were carried out from the quasi-static
region to the high dynamical regime, where the magnetic field was
pulsed at the nanosecond time scale. They have shown that the
droplet model was valid over more than 11 orders of magnitude of the
applied field sweep rate as depicted in Fig.~\ref{Moritz}. By
comparison, an attempt was made to fit the experimental data with a
SW-type barrier and it is clear that in this case the droplet model
describes nucleation processes better.

Concerning the temperature dependence of the coercive field, several
papers relate nucleation volume extrapolations using a SW barrier
type. This is usually valid in perpendicular-to-plane magnetised
system where demagnetising effects favor nucleation
\cite{PhDLemerle}. Often, the onset of magnetization reversal in the
hysteresis loops of these materials is quite abrupt, indicating that
the nucleation field is larger than the propagation field. In that
case, one can consider that the value of the coercive field is
determined by nucleation and can be used to calculate the nucleation
barrier using equations \ref{SWnucl} or \ref{VnGivord}. As
hysteresis measurements are performed in most cases in the
quasi-static regime, it is possible to consider that the barrier
height corresponds to $\Delta E_n = 25 k_BT$ and the critical field
can be deduced from this last expression.

\subsection{Nucleation centers and their
distribution}\label{defects} The models given above consider
homogeneous materials with homogeneous magnetic properties. In most
real systems, defects and non-magnetic inclusions constitute
potential nucleation centers in the case of thin film layers. Grain
boundaries, or edges for magnetic nanostructures, present a local
modification of properties such as anisotropy or exchange. Aharoni
has calculated the nucleation field for anisotropy steps or regions
with linear decrease of anisotropy and showed its strength was
reduced with the size of the defect \cite{Aharoni59,Aharoni60}.
These models qualitatively account for a distribution of nucleation
centers in real samples.

In the droplet model, it is possible to extend equation
\ref{CritBarr} when the domain wall energy is not homogeneous on the
sample. Actually exchange and anisotropy are spatially distributed
because of the granular structure. There might be also
discontinuities in their profiles due to grain boundaries. As
$\gamma$ is not constant, it leads to another expression of the
barrier height \cite{Barbara94,Moritz05}:

\begin{equation}\label{BarrDroplet}
    \Delta E_n = \frac{\pi\gamma^2t}{2\mu_0M_S}(\frac{1}{H} -
    \frac{1}{H_0})
\end{equation}

\begin{figure}[t]
\center \includegraphics*[bb= 180 401 373 570]{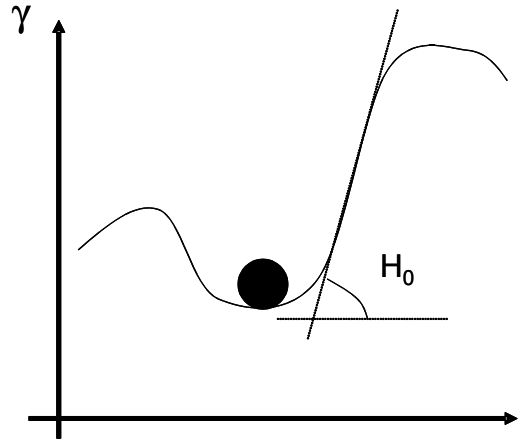}
\caption{\label{H0} Spatial fluctuations of the domain wall energy
(due to local variations of anisotropy or exchange interactions).
The critical field {\it H$_0$} is seen as the maximum slope of
$\gamma$. The black circle represents the position of the domain
wall.}
\end{figure}

The critical field {\it H$_0$} is defined as the maximum slope in
the $\gamma$ profile (see Fig.~\ref{H0}):

\begin{equation}\label{H0gamma}
    H_0 = \frac{1}{2\mu_0M_S}(\frac{\delta\gamma}{\delta r})_{max}
\end{equation}

Modeling the reversal of magnetisation by using only one nucleation
center and one nucleation barrier is most often a crude
approximation, unless the field sweep rate or temperature are very
low. Jamet {\it et al.} have extrapolated the nucleation length
(related to the nucleation volume) and the mean distance between
nucleation sites from their magnetic after-effect measurements
performed on micrometer-size dots arrays \cite{Jamet98}. The
calculated values of 26~nm and 430~nm can be seen as twice the
domain wall width and the distance separating two major defects in
the continuous layer respectively. Concerning patterned layers
approaching the domain wall width or the exchange length, it seems
that nucleation centers can be generated by the nanofabrication
itself (presence of edges for instance) or due to the extrinsic
properties of the initial continuous magnetic layer \cite{Moritz05}.
In all cases, a nucleation field or nucleation energy distribution
should be taken into account to simulate hysteresis or magnetic
after-effect measurements \cite{Bruno90,Bottoni98,Rohart06}.

The distribution of energy barriers for nucleation also accounts for
statistical effects in the magnetisation reversal. At low field
sweep rates, or at low temperature, the reversal is activated on
only a few nuclei located on major defects. If the energy barrier
for domain wall propagation is lower than the one for nucleation,
the magnetisation rapidly reverses after the first nucleation and
the hysteresis loop is quite square. In contrast, if the domain wall
propagation barrier is high, nuclei can appear on many defects
without inflating, so that the magnetisation will reverse by
nucleation. The hysteresis loop morphology is then completely
different \cite{Fatuzzo62}. Thomson and O'Grady used magnetisation
data and hysteresis loops of Tb-Fe-Co films to determine energy
barrier distributions, and the influence of this distribution on the
relative contribution of domain nucleation and domain wall
propagation to the magnetisation reversal \cite{Thomson97}. Since
domain wall propagation initiated at a limited number of nucleation
centers is a relatively slow process, a distribution of energy
barriers for nucleation in continuous films usually shows up as an
increase of the nucleation rate at increasing applied field sweep
rate \cite{Raquet96,Bruno90}. Kerr microscopy and X-ray
Photoelectron Emission Microscopy experiments performed on spin
valves, magnetic tunnel junctions and exchange bias systems
corroborate the dependence of nucleation rates on field sweep rate,
field strength and local properties
\cite{Fukumoto06,Pennec04,Romanens05,Moritz06}. These measurements
also show that due to the distribution of energy barriers,
nucleation is quite a reproducible process in most magnetic thin
films. Notable exceptions are high quality thin films that can be
grown with very few defects, like some garnets containing Rare
Earths \cite{Malozemoff79}. The influence of thermal fluctuations on
nucleation in the absence of defects has been simulated, for
instance, by Rikvold {\it et al.} \cite{Rikvold02}.

\begin{figure}[b]
\center \includegraphics*[bb= 75 412 496 563]{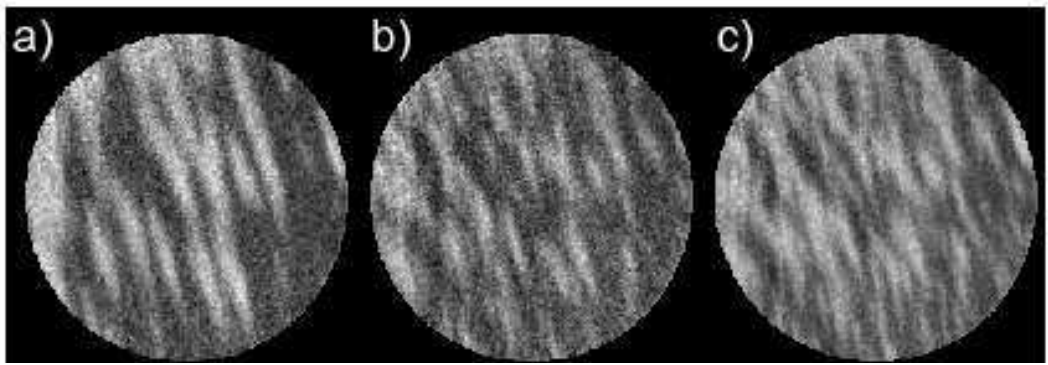}
\caption{\label{PEEMnucleation} Increase of the nucleation density
with applied field in the Fe$_{20}$Ni$_{80}$ layer of a
FeNi(4nm)/Al$_2$O$_3$(2.6nm)/Co(7nm) trilayer
\cite{Fukumoto06,Vogel05}. The images were obtained using
Photoelectron emission microscopy (PEEM) \cite{Kuch04}. The applied
fields were about 5.4, 5.8 and 6.4 mT for (a), (b) and (c)
respectively.}
\end{figure}

\section{Nucleation in bulk materials}
\label{sec-bulk} Most of the models and concepts discussed in the
previous section can also be applied to the nucleation in bulk,
3-dimensional materials. The nucleation in homogeneous bulk magnetic
materials has been considered by several authors. Broz, Braun and
coworkers \cite{Broz90,Braun91} developed a model where the
nucleation took place through the creation of a pair of Bloch or
N\'eel walls. This process is much faster than the propagation of
already existing domain walls, but in infinite bulk materials it
involves an infinite energy barrier. Aharoni and Baltensperger
\cite{Aharoni92} considered spherical and cylindrical nucleation
centers in a similar model, equivalent to the droplet model in the
previous section. They showed that in that case the energy barrier
for nucleation is still infinite at zero applied field, but will
become finite when an external magnetic field is applied.

Reversal of the magnetization in bulk materials has been
investigated especially for hard magnetic materials used in
permanent magnets. The magnetization is often hard to saturate
completely, and a number of 360$^{\circ}$ domain walls can persist
in the material \cite{Gaunt77}. Though we did not discuss this
before, this possibility also exists in 2D. When the field is
applied in the opposite direction, reversal can take place by
propagation of these existing domain walls. However, in order to
obtain a fast reversal the nucleation of new reversed domains is
necessary, since domain wall propagation is a relatively slow
process. The heterogeneous, polycrystalline microstructure of these
materials has a large influence on their magnetization reversal.
Kronm\"uller {\it et al.} analysed the temperature dependence of
$H_c$ of high coercive field permanent magnets and compared it to
the predictions of micromagnetic theories for pinning and nucleation
mechanisms \cite{Kronmuller88}. They concluded that nucleation
theory provided a better correspondance with experiment than pinning
theory, if effects of misaligned grains, local stray fields and
reduced anisotropies in grain boundaries were taken into account.
Givord {\it et al.} \cite{Givord03} developed a model where the
magnetisation reversal is associated with some critical volume
proportional to the domain wall width $\gamma$, $v(T) \propto
\gamma^3$. They derived a reversal field $\mu_0HR(T) \propto
A(T)/(v(T)^{2/3}M_S(T))$. They found that the temperature dependence
of the reversal fields in ferrite and RE-FeB (RE = Rare Earth)
magnets could be well explained by their model. The values for
anisotropy and exchange interaction in the critical volume were
close to the ones determined for the main phase. They concluded
therefore that the reversal fields were not determined by initial
nucleation at some defects.

\section{Conclusions}
Models for the nucleation of magnetisation reversal provide good
agreement with experiment results for nanoparticles. In macroscopic
systems, the onset of reversal by nucleation is in general
determined by inhomogeneities in magnetic, structural and chemical
properties, except for some very homogeneous materials with low
defect densities like ferrimagnetic garnets. For other systems,
macroscopic material parameters like exchange interaction,
spontaneous magnetisation and magnetic anisotropy can be used to
obtain an indication of nucleation fields, using different models.
However, in general a distribution of energy barriers for nucleation
exists giving rise to a dependence of the nucleation density on
temperature and applied magnetic field. In this paper, we have
discussed overcoming these energy barriers by thermal activation
and/or applying a magnetic field. Recently, evidence of quantum
nucleation by tunnel through the energy barrier has been observed in
one-dimensional single-chain magnets \cite{Wernsdorfer05}. Also, it
has been shown that spin-polarised electron currents in uniformly
magnetised layers can lead to domain nucleation \cite{Shibata05}.
These effects will probably call for new models for the nucleation
of magnetisation reversal in the coming years.


\end{document}